\documentclass[twocolumn,showpacs,preprintnumbers,amsmath]{revtex4}

\usepackage{graphicx}
\usepackage{dcolumn}
\usepackage{bm}
\begin{document}

\title{Quantum Phase Transition in the Sub-Ohmic Spin-Boson Model:  Extended Coherent-state Approach}
\author{Yu-Yu Zhang$^{1}$, Qing-Hu Chen$^{2,1,*}$, and Ke-Lin Wang$^{3,4}$}
\date{\today}

\address{$^{1}$ Department of Physics, Zhejiang University, Hangzhou 310027,
P. R. China\\
$^{2}$ Center for Statistical and Theoretical Condensed Matter
Physics, Zhejiang Normal University, Jinhua 321004, P. R. China\\
$^{3}$ Department of Physics, Southwest University of  Science and Technology, Mianyang 621010, P.  R.  China\\
$^{4}$ Department of Modern Physics, University of Science and
Technology of China, Hefei 230026, P. R. China }

\draft
\date{\today}

\begin{abstract}

We propose a general  extended coherent state approach to the qubit
(or fermion) and  multi-mode boson  coupling systems. The
application to the spin-boson model with the discretization of a
bosonic bath with arbitrary continuous spectral density is described
in detail, and very accurate solutions can be obtained. The quantum
phase transition in the nontrivial sub-Ohmic case can be located  by
the fidelity and the  order-parameter critical exponents  for the
bath exponents $s<1/2$ can be correctly given by the fidelity
susceptibility, demonstrating  the strength of the approach.
\end{abstract}

\maketitle

The spin-boson model~\cite{blume,Leg} describes a qubit (two-level
system) interacting with an infinite collection of harmonic
oscillators that models the environment acting as a dissipative
bosonic "bath". There are currently considerable interests in this
quantum many-body system due to the rich physics of quantum
criticality and decoherence~\cite{Hur,kopp,weiss}, applied to the
emerging field of quantum computations, quantum devices~\cite {mak},
and biology~\cite{reng,omer}. The  dissipative
environment~\cite{phi} in the spin-boson model is characterized by
the spectral function $J(\omega)$ with frequency behavior
$J(\omega)\propto\omega^s$. The spin-boson model undergoes a
second-order quantum phase transitions (QPT) from delocalized to
localized phase with a sub-Ohmic bath ($0<s<1$) and a
Kosterlitz-Thouless type transition in the Ohmic case ($s=1$).

To provide reliable solutions for the spin-boson model, a typical
multi-mode system, is quite challenging. Although several  numerical
methods applied to the sub-Ohmic
case\cite{Bulla,vojta1,vojta2,zheng,Wong,winter,Fehske} can
reproduce the phase diagram,  only recent quantum Monte Carlo (QMC)
simulations ~\cite{winter} and exact-diagonalization
studies~\cite{Fehske}  are capable of correctly extracting the
critical exponents in the QPT. The critical behavior in the previous
standard numerical renormalization group (NRG) calculations
\cite{Bulla,vojta1,vojta2} is incompatible with a mean-field
transition due to the failure of the quantum-to-classical mapping
with long-range interactions for $s< \frac{1}{2}$. More recently,
the early standard NRG results were improved by a modified NRG
algorithm\cite{discre}, and the mean-field behavior for $s<1/2$ was
also  reproduced.

In this paper, we present a general accurate approach to the qubit
(or fermion) and multi-mode boson coupling systems. As a important
example, we focus on the sub-ohmic spin-boson model here. It can
also be easily extended to the  famous Holstein
model\cite{THolstein} and the multi-mode Dicke model\cite{mullti}.
The crucial procedure  is to employ extended coherent states to
represent the bosonic states.  The QPT in the sub-ohmic  spin-boson
model will be analyzed  by means of the quantum information tools,
such as the ground state fidelity and fidelity susceptibility
~\cite{Zanardi,Cozzini,You}. It is a great advantage to use the
fidelity to characterize the QPT, since there should be a dramatic
change in the fidelity across the critical points. Moreover, the
non-trivial order-parameter critical exponents can be obtained with
scaling of the fidelity susceptibility.

The Hamiltonian of the spin-boson model is given by
\begin{equation}  \label{hamiltonian}
H=-\frac \Delta 2\sigma _x+\frac \epsilon 2\sigma
_z+\sum_{n}\omega_{n}a^{\dagger}_{n}a_{n}+\frac{1}{2}\sigma_{z}\sum_{n}
\lambda_{n}(a^{\dagger}_{n}+a_{n}),
\end{equation}
where $\sigma_{x}$ and $\sigma_{z}$ are Pauli matrices, $\Delta$ is
the tunneling amplitude between two levels, ${\omega_{n}}$ and
$a^{\dagger}_{n}$ are the frequency and  creation operator of the
$n$-th  harmonic oscillator, and ${\lambda_{n}}$ is the coupling
strength between the $n$-th oscillator and the local spin. The
spin-boson coupling is characterized by the spectral function,
\begin{equation}
J(\omega)=\pi\sum_{n}\lambda_{n}^{2}\delta(\omega_{n}-\omega)=2\pi\alpha
\omega^{1-s}_{c}\omega^{s}, 0<\omega<\omega_{c}
\end{equation}
with $\omega_{c}$ a cutoff frequency. The dimensionless parameter
$\alpha$ denotes the strength of the dissipation. $s=1$ stands for
an Ohmic dissipation bath. The rich physics of the quantum
dissipation is second-order QPT from delocalization to localization
for $0<s<1$, as a consequence of the competition between the
amplitude of tunneling of the spin and the effect of the dissipative
bath.

We here propose a solution of the spin-boson model by exact
diagonalization in the coherent-states space. To implement our
approach, we first perform discretization of the bath speciation
function, according to the logarithmic discretization of the
continuous spectral density $J(\omega)$ in the
NRG~\cite{Bulla,vojta1,vojta2}. The  discrete Hamiltonian is
therefore expressed as
\begin{equation}
H_{n}=-\frac \Delta 2\sigma _x+\frac \epsilon 2\sigma _z+\sum_{n}\xi
_na_n^{+}a_n+\frac{\sigma _z}{2\sqrt{\pi }}\sum_{n} \gamma _n(a_n^{+}+a_n)
\end{equation}
with
\begin{equation}
\xi _n=\gamma _n^{-2}\int_{\Lambda ^{-(n+1)}\omega _c}^{\Lambda ^{-n}\omega
_c}dxJ(x)x,\gamma _n^2=\int_{\Lambda ^{-(n+1)}\omega _c}^{\Lambda
^{-n}\omega _c}dxJ(x)
\end{equation}
In order to ensure the convergence of the results, the disctetiztion
parameter is chosen $\Lambda =2$.

The present basic scheme is similar to that in the single-mode Dicke
model~\cite{chen} and the two-site Holstein-Hubbard model~
\cite{Zhang}. For convenience, we assume that $|\varphi _1\rangle$
and $ |\varphi _2\rangle$ are the bosonic states corresponding to
spin up and down. Introducing a displacement shift parameter $
g_n=\frac{ \gamma _n}{2\xi _n\sqrt{\pi }}$~\cite{vojta2}, we propose
the following two coherent bosonic operators
\begin{eqnarray}
A_n^{+} &=&a_n^{+}+g_n,A_n=a_n+g_n \\
B_n^{+} &=&a_n^{+}-g_n,B_n=a_n-g_n.
\end{eqnarray}
The corresponding vacuum  states  $|0\rangle_{A_n}$ and $
|0\rangle_{B_n}$ are just the  coherent states in $a_n$ with
eigenvalues $\mp g_n$ in terms of $e^{\mp
g_na^{+}-g_{n}^{2}/2}|0\rangle_{a_n}$.  $|n_k\rangle_{A_k}$ and
$|n_k\rangle_{B_k}$ correspond to Fock states of the new bosonic
operators $A_k^{+}$ and $B_k^{+}$ with $n_k$ bosons for a frequency
$\omega_k$.  $|\varphi _1\rangle$ and $ |\varphi _2\rangle$ can be
expanded in the bosonic coherent states of a series of $n_k$, which
are orthonormalized in the new bosonic operators $A_k^{+} (B_k^{+})$
\begin{eqnarray}  \label{function1}
|\varphi
_1\rangle=\sum_{n_1...n_N}^{N_{tr}}c_{n_1...n_N}\prod_{k=1}^N|n_k\rangle_{A_k},
\\
|\varphi
_2\rangle=\sum_{n_1...n_N}^{N_{tr}}d_{n_1...n_N}\prod_{k=1}^N|n_k\rangle_{B_k}
\label{function2}
\end{eqnarray}
where $c'_{\{n_k\}}$ are coefficients with respect to a series of
$\{n_1,n_2...n_N\}$ for different bosonic  modes, and ${N_{tr}}$ is
the bosonic truncated number.

Then the Schr\"{o}dinger equations of the Hamiltonian (3) are
derived as
\begin{eqnarray}
-\frac \Delta 2|\varphi _2\rangle-\frac{\epsilon}{2}|\varphi
_1\rangle+\sum_{n=0}^\infty \xi _n(A_n^{+}A_n-g_n^2)|\varphi
_1\rangle &=&E|\varphi _1 \rangle \label{eq1} \\
-\frac \Delta 2|\varphi
_1\rangle+\frac{\epsilon}{2}|\varphi_2\rangle+\sum_{n=0}^\infty \xi
_n(B_n^{+}B_n-g_n^2)|\varphi _2 \rangle&=&E|\varphi _2\rangle.
\label{eq2}
\end{eqnarray}
After the substitution of Eqs. (7) and (8) and  Left multiplying the
bosonic coherent states with the both sides of Eqs. (~\ref{eq1}) and
(~\ref{eq2}),  we have
\begin{eqnarray}
\sum_{i=0}^\infty \xi _i(m_i-g_i^2)c_{\{m_k\}} &-&\frac \Delta
2\sum_{\{n_k\}}d_{\{n_k\}}\prod_{k=1}^N\;_{A_k}\langle m_k|n_k\rangle _{B_k}
\nonumber  \label{eq11} \\
&-&\frac{\epsilon}{2}c_{\{m_k\}}=Ec_{\{m_k\}}, \\
\sum_{i=0}^\infty \xi _i(m_i-g_i^2)d_{\{m_k\}} &-&\frac \Delta
2\sum_{\{n_k\}}c_{\{n_k\}}\prod_{k=1}^N\;_{B_k}\langle
m_k|n_k\rangle _{A_k}
\nonumber \\
&+&\frac{\epsilon}{2}d_{\{m_k\}}=Ed_{\{m_k\}}  \label{eq12}
\end{eqnarray}
The bosons state $|n_k\rangle$ and $|m_k\rangle$ with different
coherent bosonic operators $A_{k}^{+}$ and $B_{k}^{+}$ are not
orthogonal. The overlap can be denoted by  $ _{A_k}\langle
m_k|n_k\rangle _{B_k}=(-1)^{n_k}D_{m_kn_k} $ and $ _{B_k}\langle
m_k|n_k\rangle _{A_k}=(-1)^{m_k}D_{m_kn_k} $ with
\[
D_{m_kn_k}=e^{-2g_k^2}\sum_{i=0}^{\min \{m_k,n_k\}}(-1)^i\frac{\sqrt{m_k!n_k!%
}(2g_k)^{m_k+n_k-2i}}{i!(m_k-i)!(n_k-i)!}.
\]

According to the symmetry of Hamiltonian for $\epsilon=0$, the
coefficients  satisfy $c_{m_1...m_N}=\pm
(-1)^{\sum_km_k}d_{m_1...m_N}$. Eqs.(~\ref{eq11}) and (~\ref{eq12})
can then be transformed  into the following set of coupled equations
\begin{equation}
\mp \frac \Delta
2\sum_{n_1...n_N}c_{\{n_k\}}\prod_{k=1}D_{m_kn_k}+\sum_{i=0}^\infty \xi
_i(m_i-g_i^2)c_{\{m_k\}}=Ec_{\{m_k\}}  \label{equation}
\end{equation}

A complete implementation of the numerical diagonalization is
described below to obtain the amplitudes set of $c_{\{n_k\}}$ of the
bosonic  state $\varphi_1$ ($\varphi_2$). The Hilbert space can be
labeled by a vector $\overrightarrow{n}=(n_1,...,n_N)$ with $n_k=0,
1,..., N_{tr}$. The sum of bosonic number $n_k$ is restricted to
truncated number $N_{tr}$, e.g. $\sum n_k\leq N_{tr}$. For example,
with a set of $N=3, N_{tr}=3$ the involved configurations of bosonic
states $|n_1,n_2,n_3>$ are expressed as following:
\[
|000\rangle,
\]
\[
|100\rangle,|010\rangle,|001\rangle,
\]
\[
|200\rangle,|110\rangle,|020\rangle,|101\rangle,|011\rangle,|002\rangle,
\]
\[
|300\rangle,|210\rangle,|120\rangle,|030\rangle,|201\rangle,|111\rangle,
|021\rangle,|102\rangle,|012\rangle,|003\rangle.
\]
Consequently, the total number of basis states $N_s=20$. To obtain
the true exact results, in principle, the number of bosonic modes $N
$ and the truncated number $N_{tr}$ should be taken to infinity.
Fortunately, in the present calculation, setting  $N=16$, which is
big enough in  NRG, and $N_{tr}=5$ is sufficient to give very
accurate results with relative errors less than $~10^{-5}$ in the
whole parameter space. The following results are just obtained with
$N=16$ and $N_{tr}=5$.

To study the QPT in the sub-Ohmic spin-boson model, we employ the
ground state fidelity to locate the critical point $\alpha_c$. A
simple expression of the ground-state fidelity is given just by the
modulus of the overlap
\begin{eqnarray}
F(\alpha,\alpha^{\prime})=|\langle\varphi_{1}(\alpha)|\varphi_{1}(\alpha^{
\prime})\rangle+
\langle\varphi_{2}(\alpha)|\varphi_{2}(\alpha^{\prime})\rangle |
\end{eqnarray}
The QPT is expected to be signaled by a drop in the fidelity
corresponding to two arbitrarily neighboring Hamiltonian parameters
$\alpha^{\prime}=\alpha+\delta\alpha$~\cite {Zanardi,You}. Based on
the normalized ground states $|\varphi_1\rangle$ and
$|\varphi_2\rangle$, we now illustrate our results obtained by
numerically diagonalization of Eq.(~\ref{equation}).

\begin{figure}[tbp]
\includegraphics[width=9cm]{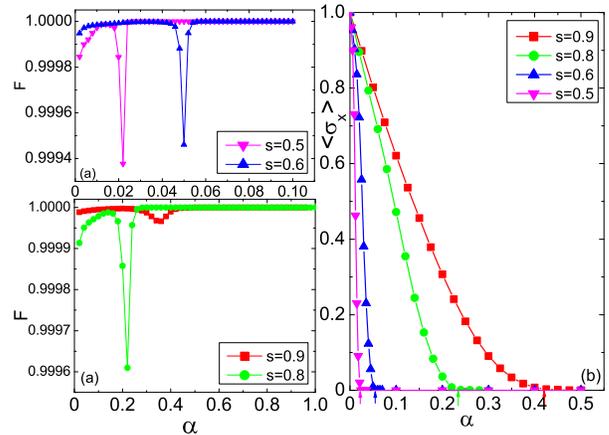}
\caption{  (Color online). (a) The fidelity $F$ as a function of
$\alpha$ for various values of  $ s$ and  (b) The tunneling
$\langle\sigma_{x}\rangle$ between two states of the spin as a
function of $\alpha$ in the case of $\Delta=0.01$ and $\epsilon=0$..
} \label{fidcrit}
\end{figure}

Fig. ~\ref{fidcrit} (a) shows the behavior the fidelity
$F(\alpha,\alpha^{\prime})$ in the sub-Ohmic case with
$s=0.5,0.6,0.8,0.9$ for the spin tunneling amplitude $\Delta=0.01$ .
A sharp drop at the critical point $\alpha_{c}$ separates the
delocalized phase at small $\alpha$ and the localized phase at large
$\alpha$. So it is evident that  we can locate the critical points
$\alpha_c$ efficiently by the ground state fidelity. It is
interesting that the ground state fidelity does not drop to $0$ at
critical points, demonstrating a continuous QPT\cite{first-order}.

The QPT from delocalized to localized phases can also be shown by
behavior of  the tunneling motion $\langle\sigma_{x}\rangle$ between
spin up and spin down \cite{zheng}. $\langle\sigma_{x}\rangle$ goes
rapidly to zero in the localized phase and is finite in the
delocalized phase. The $\alpha(\Delta,s)$ dependence of
$\langle\sigma_{x}\rangle$ is shown in Fig.~\ref {fidcrit}(b). For
all value of $s$, we observe that $\langle\sigma_x \rangle$ is
continuous at the transition. The discontinuous behavior observed
previously \cite{zheng} may be attributed to the special variational
approach itself, and is perhaps worthy of a further study.

\begin{figure}[tbp]
\includegraphics[width=8cm]{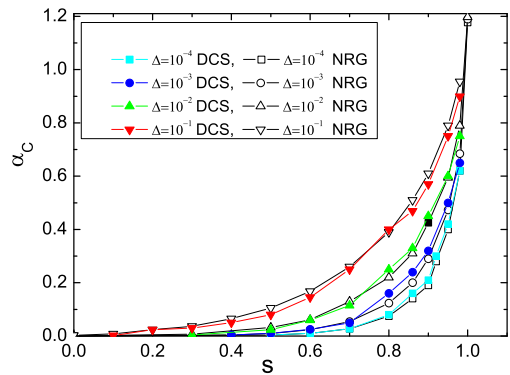}
\caption{ (Color online). The delocalized-localized transition point
$\alpha_{c}$ as functions of $s$ obtained by the present approach in
the case $\Delta=10^{-1},10^{-2},10^{-3},10^{-4}$. The NRG data are
also shown for comparison.} \label{alpha}
\end{figure}

As discussed above, both the fidelity $F(\alpha,\alpha^{\prime})$
and tunneling parameter $\langle\sigma_{x}\rangle$ can be used to
locate the critical points of the QPT. We observe that both
quantities can give nearly the same  critical points. The phase
boundaries obtained by either methods as a function of $s$  is
plotted in Fig.~\ref{alpha} in case of the tunnel splitting $\Delta$
ranging from $ 10^{-4}$ to $10^{-1}$ for $\epsilon=0$. The results
from previous NRG techniques\cite{Bulla,vojta1} are also collected
for comparison. It is interesting to note that the present results
for the critical points are in good agreement with the NRG ones.
Because we also use the truncated NRG Hamiltonian (3), the critical
points should be slightly above the QMC ones\cite{winter} where all
frequencies are included. For fixed truncated  Hamiltonian , as
$N_{tr}$ increases, $\alpha_c$ converges to the true value from
above very quickly. We believe that we obtain the converging
critical points for any values of $\Delta$ for fixed $N$ in the
present work.

The ground-state fidelity susceptibility $\chi$ is defined as the
second derivative of the fidelity~\cite{Zanardi,You}
\begin{equation}
\chi(\alpha)=2\lim_{\delta \alpha\rightarrow0}\frac{1-F(\alpha,\delta
\alpha) }{\delta \alpha^{2}}.
\end{equation}
Since $\chi$ is independent of the arbitrary small parameter $\delta
\alpha$ , it is regarded as a more effective tool to detect the
singularity in QPT. As addressed in Ref.~\cite{You}, the fidelity
susceptibility is similar to the magnetic susceptibility. In the
localized phase, the scaling behavior of fidelity susceptibility
$\chi$ obeys
\begin{equation}
\chi(\alpha)\propto|\alpha-\alpha_c|^\beta
\end{equation}

\begin{figure}[tbp]
\includegraphics[width=9cm]{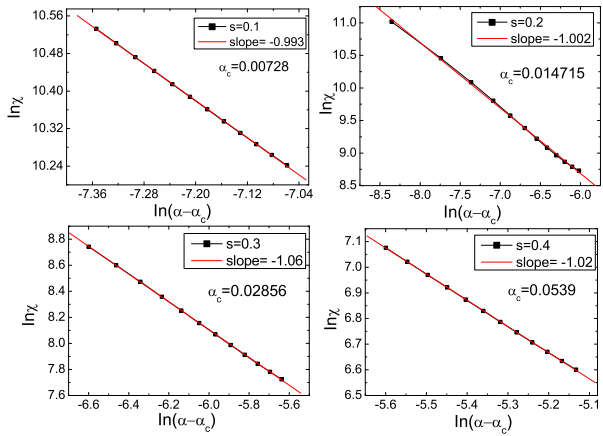}
\caption{ (Color online). The scaling behavior of the fidelity
susceptibility $ \chi\sim|\alpha-\alpha_{c}|^{\beta}$ in log-log
scale for $s=0.1,0.2,0.3,0.4$ with $\Delta=0.1$. The data fit well
with the straight line with slope very close to $-1$.}
\label{fidscal}
\end{figure}

Recently, the studies from the QMC approach\cite{winter}, the
exact-diagonalization~\cite{Fehske}, and the modified
NRG\cite{discre} have shown that the quantum-to-classical mapping is
valid for the sub-Ohmic spin-boson model, i.e. the critical
exponents are classical, mean-field like, in contrast with the early
standard  NRG calculations where the quantum-to-classical mapping is
suggested to fail for $s<1/2$. It was argued ~\cite{winter} that the
standard NRG is not able to capture the correct physics in the
localized phase. It is known that the localized phase is two-fold
degenerate. Our ansatz Eqs. (7-10) is just proposed for these two
states, and the unknown coefficients can be obtained by solving the
Schr\"{o}dinger equations  very accurately. Therefor, we will
extract the susceptibility critical exponent  for $s<1/2$, to
address this crucial controversy.

Fig.~\ref{fidscal} presents that the fidelity susceptibility $ \chi$
as a function  $(\alpha-\alpha_c)$ for $s=0.1,0.2,0.3,0.4$ in
log-log scale. All curves show almost perfect straight line with a
slope very close to $-1$, demonstrating that the susceptibility
critical exponent may be just equals to $\beta=-1$ with
$\chi\sim(\alpha-\alpha_c)^{-1}$. Recently, the critical exponent of
the magnetic susceptibility has been estimated to be $-1$ by QMC
simulations ~\cite{winter} and exact-diagonalization
studies~\cite{Fehske}. We do not think this is a coincidence. You et
al~\cite{You} have shown a neat connection between the fidelity
susceptibility $ \chi$ and the magnetic susceptibility $ \chi_m$
through $ \chi=\chi_m/4K_B T$ ($K_B$ is the Boltzmann constant, $T$
is the temperature). We believe this relation is also applicable to
zero temperature, and therefore these two susceptibilities can give
the same order-parameter critical exponents in QPT. We also confirm
the mean-field behavior for $1/2<s<1$.

In summary, we have introduced an efficient algorithm in the new
bosonic coherent Hilbert space and presented reliable solution for
the sub-Ohmic spin-boson model. The ground-state fidelity, which is
a quantum information tool, is employed to locate the critical
coupling strength $\alpha_c$ of the QPT . The transition from the
localized phase to delocalized phase is accompanied by a minimum of
the fidelity. Furthermore, the fidelity susceptibility gives the
order-parameter critical exponent $\beta=-1$ in the case $s<1/2$,
which agrees well with the exponent of magnetic susceptibility. Both
behaviors of  the tunneling $ \langle\sigma_x\rangle$  and the
fidelity around the critical point  exclude the possibility of the
first-order QPT. We stress that all eigenstates and eigenvalues of
the spin-boson model can  be obtained accurately and many observable
can be calculated directly within the present approach. The present
technique to deal with bosons would be combined with other
established methods.

We thank Tao Liu and Ninghua Tong for useful discussions and
especially  Ninghua Tong for  providing the data of NRG. This work
was supported by National Natural Science Foundation of China,
PCSIRT (Grant No. IRT0754) in University in China, National Basic
Research Program of China (Grant No. 2009CB929104), Zhejiang
Provincial Natural Science Foundation under Grant No. Z7080203, and
Program for Innovative Research Team in Zhejiang Normal University.

$*$ Corresponding author. Email:qhchen@zju.edu.cn


\begin{references}
\bibitem{first-order} In the case of the first-order QPT, the wave functions at two sides of the critical
points  are orthogonal due to  the level crossing,  the  fidelity
should drops to $0$ at QPT.
\end{references}
\end{document}